\documentclass[pdflatex]{revtex4}
\pdfoutput=1

\usepackage{graphicx}
\usepackage{dcolumn}
\usepackage{bm}

\begin{document}

\title{Self-Feeding Turbulent Magnetic Reconnection on Macroscopic Scales}

\author{Giovanni Lapenta}
\affiliation{Centrum voor Plasma-Astrofysica, Departement Wiskunde,
Katholieke Universiteit Leuven, Celestijnenlaan 200B, B-3001 Leuven,
Belgium.}

\date{\today}

\begin{abstract}
Within a  MHD approach we find magnetic reconnection to progress in
two entirely different ways. The first is well-known: the laminar
Sweet-Parker  process. But a second, completely different and
chaotic reconnection process is possible. This regime has properties
of immediate practical relevance: i) it is much faster, developing
on scales of the order of the Alfv\'en time, and ii) the areas of
reconnection become distributed chaotically over a macroscopic
region. The onset of the faster process is the formation of closed
circulation patterns where the jets going out of the reconnection
regions turn around and forces their way back in, carrying along
copious amounts of magnetic flux.
\end{abstract}

\pacs{52.35.Vd,52.30.Cv,52.30.-q}
\maketitle

Reconnection is one of the most active areas of research in plasma
physics~\cite{biskamp-book,priest-forbes}. Reconnection is believed
to be a crucial engine of energy conversion in astrophysical objects
such as the environment of black holes~\cite{kronberg} and
stars~\cite{priest-forbes} and in laboratory experiments~\cite{mrx}.
In reconnection, magnetic field lines break and reconnect changing
their topological connectivity~\cite{biskamp-book} and in the
process convert magnetic energy into kinetic and thermal energy.
Explaining how reconnection can be an active agent for energy
exchanges in macroscopic systems require to address two fundamental
problems.

The first problem is that the detailed study of reconnection leads
to the conclusion that reconnection is a very localized process
developing in tiny regions (called diffusion regions) within the
overall system. However effective such a localized process may be,
still how can it affect large fractions of the system energy? There
is a need to explain how a vast area of magnetic field and energy
can undergo such a process when it takes place on very small scales.
A suggestion~\cite{matthaeus:2513,drake} has been made that
reconnection might take place in large areas in the form of a
cluster of many diffusion regions filling a significant area of the
domain. This proposal is very attractive, but evidence for such a
process is still lacking, either as direct observational evidence or
as simulation demonstration.

A second difficulty is to achieve the required rate of reconnection.
Reconnection requires dissipative processes usually not present in a
simple description of the plasma as a resistive fluid: the level of
resistivity present in the system is vastly insufficient to explain
the observed rates. Reconnection can be fast on the microscopic
scales~\cite{birnGEM} or when the process of reconnection is driven
by flows~\cite{biskamp-book} (spontaneously generated in the system
or created externally).

We report here a possible mechanism capable of inducing a turbulent (meant here simply to imply a chaotic process)
reconnection region encompassing a large scale portion of  a
macroscopic system and where reconnection  aliments itself requiring
no external flows to keep a fast rate of reconnection.

{\it Reference systems considered:} For simplicity we consider two
types of systems initially in a 1D equilibrium state described
either as a balance of magnetic and plasma pressure (the so-called
Harris sheet):
\begin{equation}
{\bf B}(z)=B_0 \{\tanh(z/L),0,0\}\, ; \; p(z)=p_0  \,{\rm sech}(z/L)
\label{harris}
\end{equation}
or as a force free equilibrium in a uniform plasma:
\begin{equation}
{\bf B}(z)=B_0 \{\tanh(z/L),{\rm sech}(z/L),0\}\, ; \; p(z)=p_0
\label{force-free}
\end{equation}

To follow a now standard procedure that facilitates comparison with
previously published works, the evolution of the system is initiated
by an initial perturbation  chosen according to the so-called GEM
challenge~\cite{birnGEM}: $\delta A_y=\epsilon B_0 L \cos(2\pi
(x-L_x/2)/L_x) \cos(\pi z/L_z)$, with $\epsilon=0.1$. We consider
the 2D plane $(x,z)\in[0, L_x]\times [-Lz/2, L_z/2]$ where
reconnection develops.

The aim of the present paper is to consider macroscopic processes
(on scales much larger than the ion inertial length), therefore
 the fluid MHD approach is appropriate, compared with the kinetic
approach valid at all scales but relevant only at small scales below
what is considered here. We use the FLIP3D-MHD
code~\cite{jub-flipmhd}, based on  the visco-resistive MHD equations
and including an energy equation and and ideal equation of state
with adiabatic index  $\gamma=5/3$. The simulations  reported below
have different system sizes (listed  in each case) but all have grid
spacing $\Delta x/L, \Delta z/L=1/12$ and  time step  $\Delta
t/\tau_A=.05$ (with the Alfv\'en time $\tau_A=L/v_A$). This level of
accuracy results in converged solutions, as tested by comparing
simulations with a time step or grid spacing increased separately by
a factor of two. Periodic boundary conditions are used along $x$ to
try to mimic similar events happening in nearby regions of the
system, as proposed in the mechanism suggested in Ref.~\cite{drake}.
In $z$ the so-called no-slip conditions (i.e. no parallel flow is
allowed and the boundary is impermeable to the plasma; the magnetic
field remains parallel to the wall) are used. This choice of the
boundary conditions could impede the flow patterns that will be
analyzed below, reducing the rate of reconnection compared with an
open system). For this reason we have conducted simulations with
varying vertical box size ($L_z=40, 60, 80, 100, 120$) and we have
compared with open boundary conditions (as in
Ref.~\cite{lapenta-knoll-apj}). The results are not affected,
qualitatively or quantitatively, proving that  the boundary in $z$
is far enough from the reconnecting layer as not to affect the
evolution. Viscosity and resistivity are uniform and are expressed
via the Reynolds ($R$) and Lundquist ($S$)
numbers~\cite{biskamp-book}.

The evolution of the topology of the magnetic surfaces and of the
stream function is monitored. The intersection of the magnetic
surfaces in the plane of a  2D system are easily obtained as contour
lines of the component of the vector potential along the ignorable
direction ($y$ in our choice of coordinate system):
$A_y$~\cite{biskamp-book}. The stream function $\psi$ is a fluid
quantity that plays a similar role for the streamlines that $A_y$
plays for the magnetic surfaces~\cite{biskamp-book}.

\begin{figure}
\centering
\begin{tabular}{c}
a)\\
\includegraphics[width=80mm,height=40mm,angle=0]{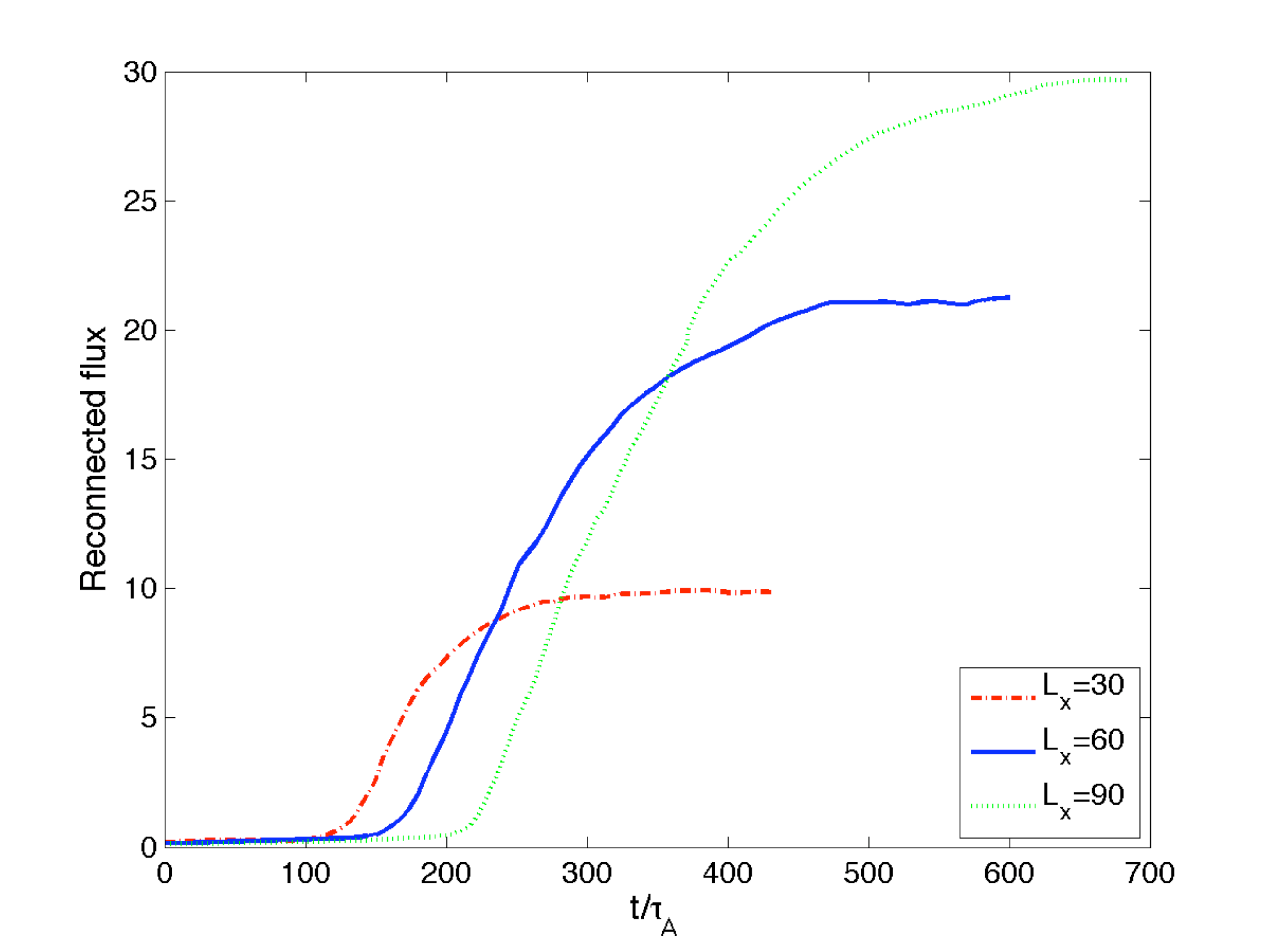}\\
b)\\
\includegraphics[width=80mm,height=40mm,angle=0]{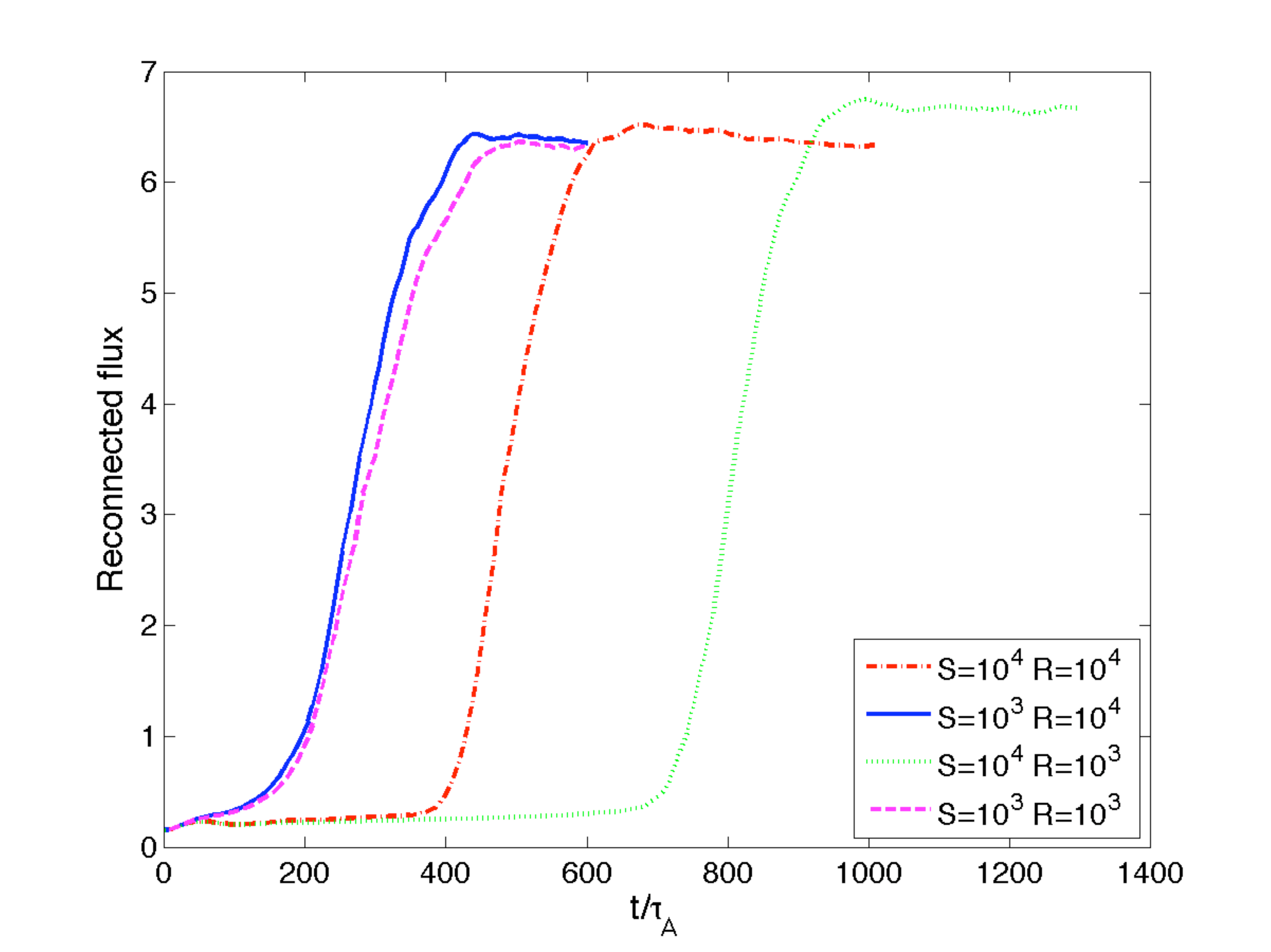}
\end{tabular}\caption{Reconnection rate for two types of equilibria. Note the different vertical and
horizontal scales in panel a and b. Panel a: Harris sheet,
eq.~(\ref{harris}). Different horizontal system sizes are used:
$L_x/L=30, 60, 90$, with the same vertical size  ($L_z/L=120$),
viscosity (Reynolds number, $R=10^4$) and resistivity (Lundquist
number, $S=10^4$). Panel b: Force free, eq.~(\ref{force-free}).
Different viscosity (Reynolds number, $R$) and resistivity
(Lundquist number, $S$) are used: $R=10^3,10^4$ and $S=10^3,10^4$,
for the same system size: $L_x/L=30$, $L_z/L=40$. }\label{rate}
\end{figure}

{\it Two stage system evolution:} The system evolves in two phases.
Figure~\ref{rate} shows the evolution of the reconnected flux (i.e.
the amount of magnetic flux that has passed through the reconnection
process) from its initial configuration with only open field lines
starting on one vertical boundary and exiting the other to its final
state with a new topology including field lines connected to the
same vertical boundary at both ends (referred to as closed, see
Fig.\ref{circulation} below). The reconnected flux is eventually
collected towards the two ends of the system ($x=0,L_x$) of the
system because of the choice of periodic boundary conditions and of
the initial perturbation (symmetric and strongest in the center).

The reconnected flux (that causes the presence of the new closed
field lines) is measured as described in the textbook procedure for
2D systems~\cite{biskamp-book}: the out of plane vector potential
$A_y$ is computed and on the mid axis ($z=0$) its maximum and
minimum are computed, their difference, initially zero, provides the
amount of flux in the closed field lines, caused by reconnection.

To show the robustness of the processes discussed, different
equilibria and dissipations are used. For both types of equilibria,
there is one slow reconnection phase followed by a vastly faster
process where almost all reconnection happens.

The evolution shown in Fig.~\ref{rate} presents clearly two phases.
At first, the flux is very slowly reconnected and the details of the
the growth (slope of the curve) and its duration vary widely with
viscosity and resistivity (measured by $R$ and $S$). The second
phase is much faster and is rather insensitive to both viscosity
and resistivity.

We focus here on the transition between the slow phase and the fast
phase showing that it is linked to the formation of a self-feeding
process where the fast flow out of the reconnection regions is
recycled into the inlet of the reconnection region causing a
feedback loop where the reconnection process feeds on itself.
Furthermore we show that the process is chaotic leading to multiple
short lived reconnection regions popping up randomly, frequently and
at multiple locations simultaneously. Even though the process is
dynamical and active with a continuous creation and destruction of
reconnection sites, the overall rate remains remarkably steady: the
reconnected flux increases monotonically during the fast
reconnection phase shown in Fig.~\ref{rate}, despite the dynamical
physics behind it.

Each phase needs resistivity as its core dissipative mechanism: in a
complete kinetic description this feature would change the details
of the rate of reconnection in each phase but the overall transition
between the two phases is not a process dependent on the presence of
resistivity or on its value. Similarly viscosity is not not a key
element. Let us turn the attention now on this mechanism allowing
the transition from slow to fast reconnection.

{\it Self-feeding and the nature of faster reconnection:} The
transition to the fast reconnection process is characterized by the
transition from a state where the outflow from a single reconnection
region remains localized near the central horizontal axis of the
system  to a state where the outflow spills into the bulk of the
system and forms a circulation loop out of the reconnection region
and directly into it. Figure \ref{circulation} shows the plasma
circulation at different times, as measured by the stream function
of the plasma. At the early time during the slow reconnection phase,
the jets from the one and only reconnection site caused by the
initial perturbation remains bound to the axis in the outflow region. At the later time
during fast reconnection, loops are formed in the plasma flow that link the outflow with
the inflow to the reconnection region.
The fast reconnection process is accompanied by a direct
circulation pattern between the inlet and outlet of the reconnection
sites. It is as if a pipe was feeding the plasma exiting at fast
speeds from the reconnection region back into the inflow: but at each
passage, the flow brings in new magnetic field lines that become decoupled with the plasma
in the reconnection region and contribute to
the global reconnection on large scales.

\begin{figure}
\centering
\begin{tabular}{c}
a)\\
\includegraphics[width=80mm,height=50mm,angle=0]{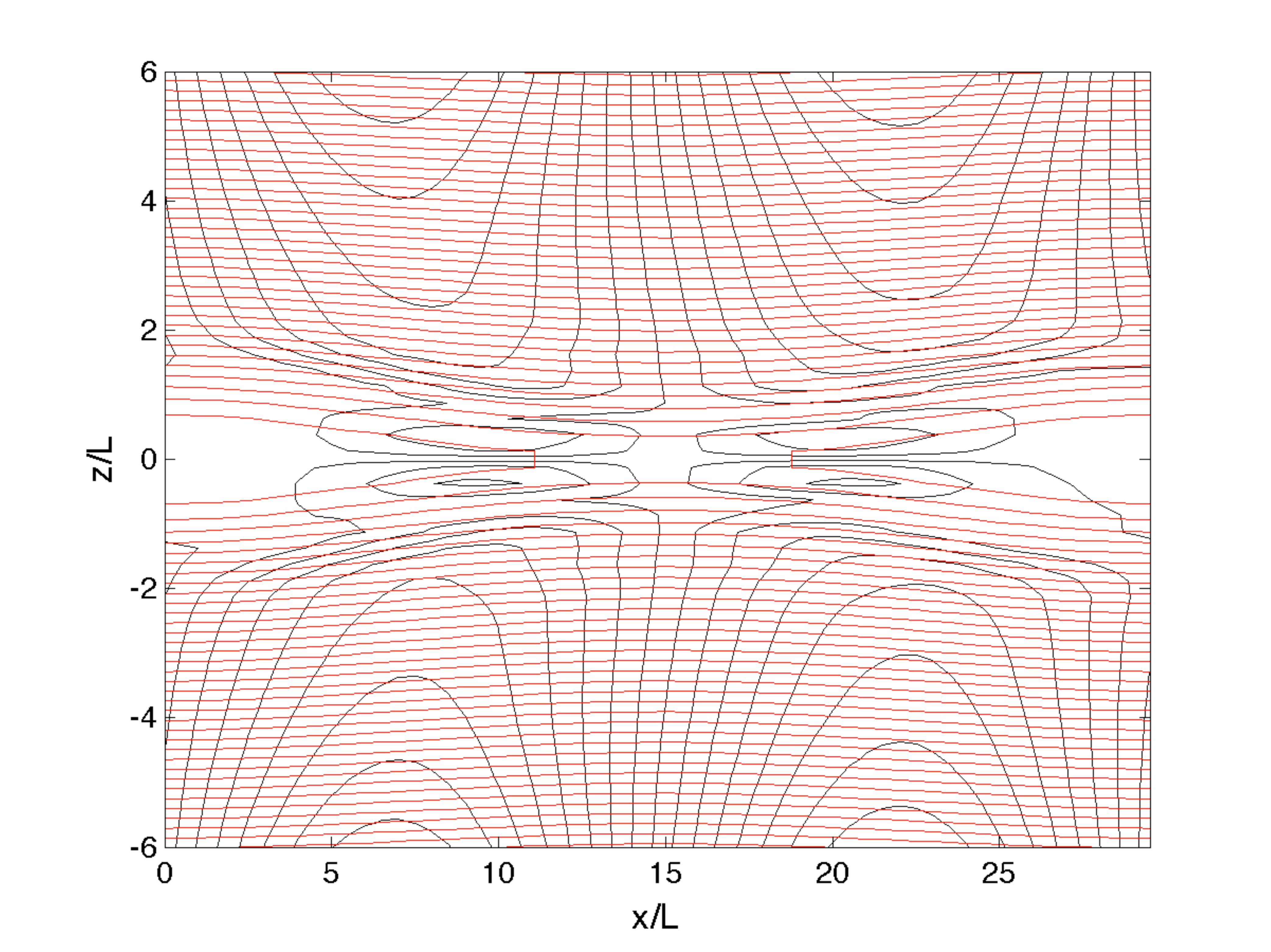}\\ b)\\
\includegraphics[width=80mm,height=50mm,angle=0]{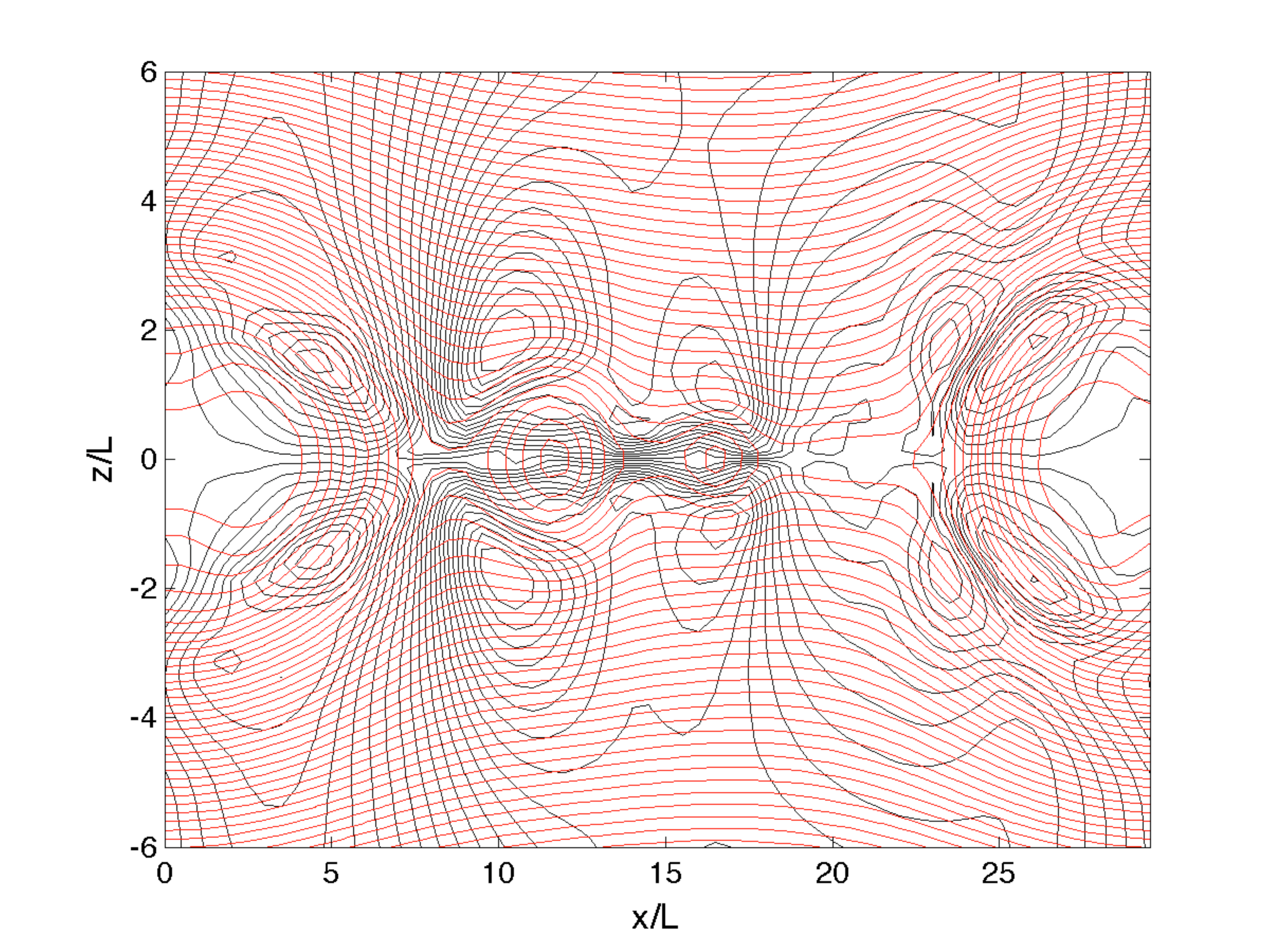}\\
\end{tabular}\caption{Magnetic topology and flow pattern at two times,
during the slow phase (a: $t/\tau_A=100$) and during the fast phase
(b: $t/\tau_A=200$). In red: intersection of the magnetic surfaces
with the plane of the simulation. In black:  contours of the stream
function, corresponding to the circulation lines  everywhere tangent
to the flow speed. Results from the Harris equilibrium run with
$L_x/L=60$, $L_z/L=40$, $R=10^4$ and $S=10^4$. Blow up around the
central axis.}\label{circulation}
\end{figure}

In the fast reconnection phase, the reconnection process changes
nature. The Sweet-Parker (SP) layer~\cite{priest-forbes} present
during the slow phase of reconnection becomes destabilized and
multiple islands form. In between islands x-points form where each
reconnection site is driven by its own self-feeding circulation
pattern (as well as by neighboring reconnection
sites~\cite{parnell}).  In the fast phase, the reconnection process
resembles more the x-point configuration of driven reconnection than
the y-point configuration of spontaneous SP
reconnection~\cite{priest-forbes}, thereby enabling the faster rate.

In smaller systems (smaller than reported here), the process
described above is inhibited by the limited size in the horizontal
 direction. Under those circumstances, the SP layer may
become unstable to secondary islands (as reported in
Ref.~\cite{birn-hesse} for the so-called GEM challenge) but the
transition to the chaotic stage of reconnection is possible only in
large macroscopic systems, as shown above.

\begin{figure}
\centering
\includegraphics[width=80mm,height=50mm,angle=0]{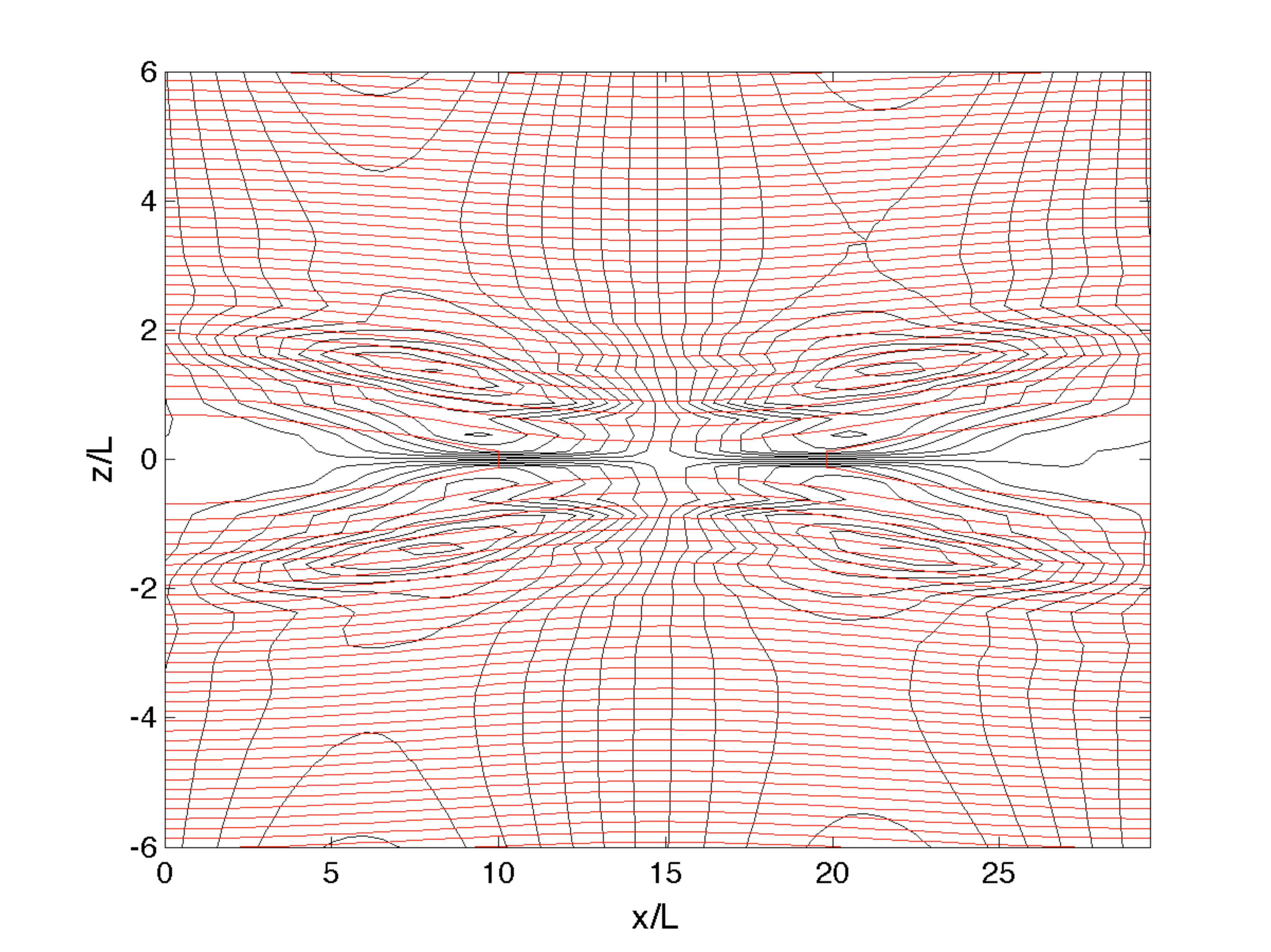}
\caption{Magnetic topology and flow pattern shown for the same case
as in Fig.~\ref{circulation} but at a time right before the start of
the fast phase ($t/\tau_A=150$). }\label{circulation2}
\end{figure}

The transition between slow and fast reconnection is linked to the
formation of these self-feeding circulation patterns: during the
slow phase the flows have not yet formed  the self-feeding patterns.
The results support the view that the formation of the circulation
pattern precedes the onset of faster reconnection.
Figure~\ref{circulation2} shows the magnetic topology and flow
pattern at a time when reconnection is still progressing slowly.
Clearly the pattern is forming. The observation of subsequent frames
(visible in movies) shows the first formation of the self-feeding
loops during the slow phase, its subsequent strengthening until
eventually the reconnection rate takes off strongly. There is a time
interval when even though the speed of reconnection has not
increased yet, the circulation pattern is already forming, thereby
preceding the onset of faster reconnection. This fact suggests (but
does not prove) that  the circulation pattern is indeed the cause of
fast reconnection and not one of its effects.

The destabilization of the outflowing jets from the laminar
reconnection phase and the formation of the self-feeding loops is
due to the interaction of the newly reconnected plasma with the
magnetic flux already accumulated in the outflow region. This
process is more effective for smaller systems where less room is
available for the reconnected  flux. Indeed, as $L_x/L$ is increased
from  30 to 60 to 90 the time to onset increases linearly.
Furthermore, the destabilization of the flows and the onset of
faster reconnection is delayed  if the outflow is impeded by
dissipations (lower $S$ and $R$)  or by the presence of an out of
plane magnetic field (as in the case of the initial force-free
equilibrium) that diverts part of the kinetic energy gained in the
reconnection region towards the out of plane direction.

{\it Multiplicity of  reconnection sites:} The self-feeding process
described above is not steady: the islands are continuously created
and destroyed in a chaotic process. Figure \ref{circulation}-b shows
a  pattern of different reconnection regions in proximity of
different islands and self-feeding circulation patterns each with
its own size, some are emerging others are dying off to be replaced
continuously by new ones. For comparison,   the slow phase of
reconnection (see Fig.~\ref{circulation}-a) has just one single long
SP layer, a well known feature of slow laminar
reconnection~\cite{priest-forbes}.

\begin{figure}
\centering
\includegraphics[width=80mm,height=60mm,angle=0]{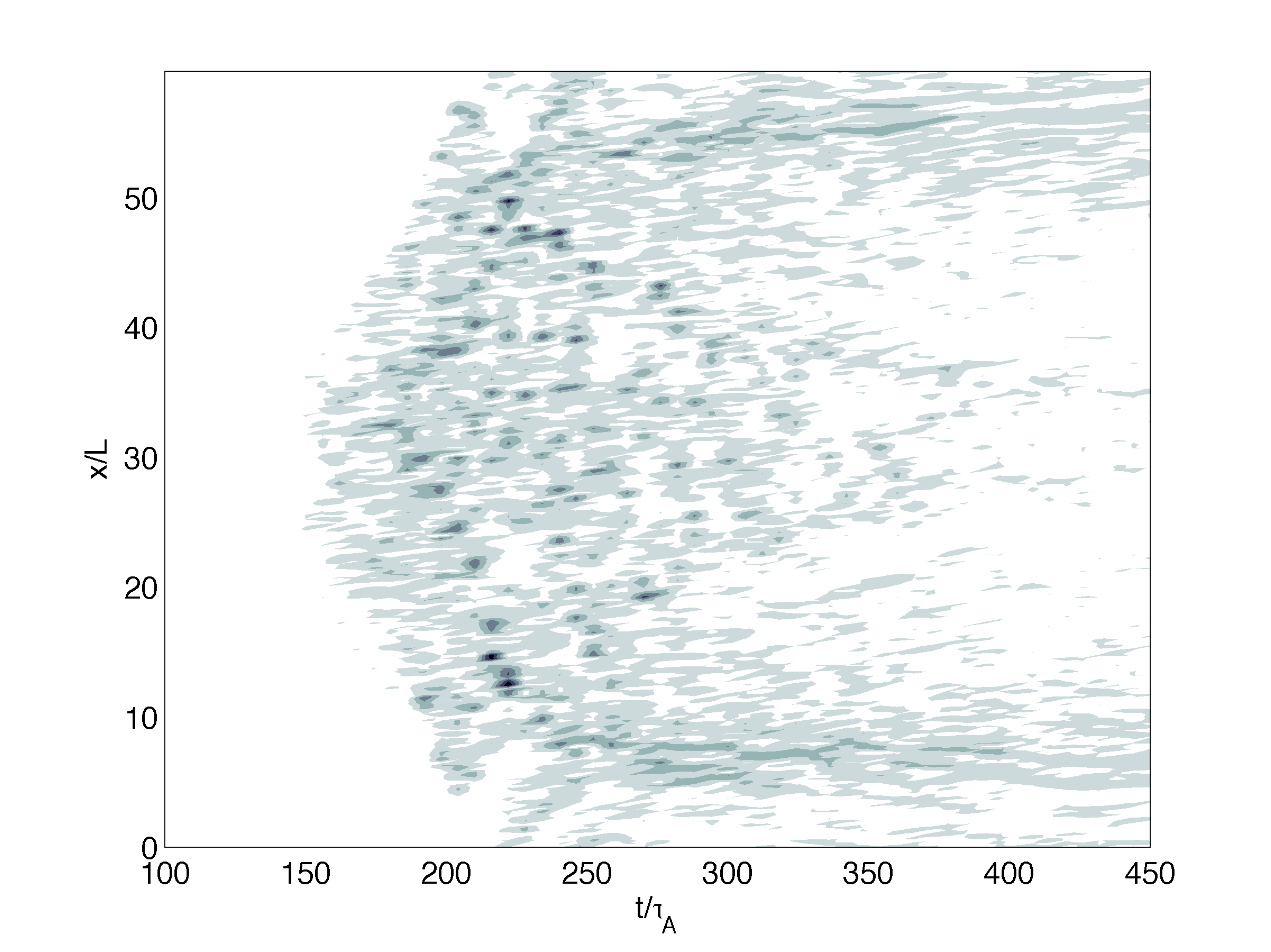}
\caption{Evolution in space (vertical axis) and time (horizontal
axis) of the magnetic islands. Results from the Harris run with
$L_x/L=60$, $L_z/L=40$, $R=10^4$ and $S=10^4$. Note that a limited
temporal span is shown.}\label{turbulence}
\end{figure}

Figure \ref{turbulence} shows the space-time evolution of the
magnetic islands in the central horizontal plane of the simulation
with $L_x=60$. Magnetic islands correspond to regions of increased
curvature of the out of plane component of the vector potential:
darker regions in the plot correspond to regions of increased
$|\partial^2 A_y/\partial x^2|$. Other typical indicators of
reconnection  (e.g. reconnection current or reconnection electric
field) lead to similar plots, not reported. Different reconnection
regions and different islands are continuously created with a
limited life span. At any given time, there is the contemporary
presence of multiple reconnection sites. The reconnection process is
very dynamical and chaotic, even though the overall accounting of
the amount of flux processed by reconnection progresses steadily (as
shown in Fig.~\ref{rate}).

We remark two crucial differences of the chaotic reconnection
region, compared to previously considered turbulent scenarios.
First, the transition towards a turbulent reconnection process is
spontaneous and it is not initiated by imposing turbulent fields or
flows~\cite{fan:5605}. Second, the chaos considered here is
intrinsic of the fluid model and bears no relationship with the
microscale (kinetic) turbulence embodied by the so-called anomalous
resistivity: here resistivity is fixed and uniform, the independence
of the rate of reconnection from resistivity is caused by the
mechanism produced by the self-feeding closed circulation loops.

{\it Recapitulation:} Results are reported above using two different
types of equilibria:  Harris  and force free equilibria. In both
cases the system given a standard initial perturbation goes through
two stages. The first is the well known SP laminar
reconnection~\cite{priest-forbes} that crawls on the resistive time
scales. The second phase takes off at later times, depending on the
parameters and the initial equilibrium, and corresponds to a faster
and turbulent reconnection process. The chaotic stage is anticipated
by the destabilization of the outflowing reconnection jets which
turn back towards the reconnection region and form a conveyor-belt
closed-circulation loop that carries quickly new magnetic flux at
the reconnection point where it becomes decoupled from the flow and
contributes to the overall macroscopic reconnection process. The
flow pattern and the current layer become chaotic with recurrent
changes of number and locations of reconnecting points and magnetic
islands. The process of reconnection becomes fast as it is driven by
the incoming flow due to the self-feeding of the closed circulation
patterns.

{\it \small The author is grateful to R. Keppens for testing one of
the simulations above with AMRVAC. Work  supported by the {\it
Onderzoekfonds K.U. Leuven} and by the EC via the SOLAIRE network
(MRTN-CT-2006-035484).
Simulations conducted on the HPC cluster VIC of the K.U. Leuven.}

\end{document}